\documentstyle[12pt]{article}

\begin{document}
\begin{titlepage}
\title{EXTREMELY CHARGED STATIC DUST DISTRIBUTIONS IN GENERAL RELATIVITY}
\vspace{5cm}
\author{MET{\' I}N G{\" U}RSES\\
{\small Mathematics department, Bilkent University},\\
{\small 06533 Ankara-TURKEY} \\
E-mail: gurses@fen.bilkent.edu.tr}
\maketitle

\noindent
A talk given in the {\it International Seminar on Mathematical
Cosmology}, Potsdam-Germany, March 30- April 4, 1998. To appear
in the proceedings, Eds. H.-J. Schmidt and M. Rainer, World Scientific,
Singapore

\begin{abstract}
Conformo static charged dust distributions
are investigated in the framework of general relativity.
Einstein's equations reduce to a nonlinear version of Poisson's
equation and Maxwell's equations imply the equality of the charge
and mass densities. An interior solution to the extreme
Reissner-Nordstr{\" o}m metric is given. Dust distributions
concentrated on regular surfaces
are discussed and a complete solution is given for a spherical thin shell.
\end{abstract}

\end{titlepage}

\section{Introduction}

Let $M$ be a four dimensional spacetime with the metric

\begin{equation}
g_{\mu\, \nu}=f^{-1}\,\eta_{1\, \mu\, \nu}-u_{\mu}\,u_{\nu} \label{mt1}
\end{equation}

\noindent
where $\eta_{1\, \mu\, \nu}=$ \,diag $(0,1,1,1)$ and $u_{\mu}=\sqrt{f}\,
\delta_{\mu}^{0}$.
Here Latin letters represent the space indices and $\delta_{ij}$ is the
three dimensional Kronecker delta. In this work we shall use the same
convention as in \cite{hwk}. The only difference is that we use Greek
letters for four dimensional indices. Here $M$ is static.
The inverse metric is given by

\begin{equation}
g^{\mu \, \nu}= f\, \eta_{2\,}^{ \mu \, \nu}-u^{\mu}\, u^{\nu} \label{mt2}
\end{equation}

\noindent
where $\eta_{2\, \mu\, \nu}=$ \,diag $(0,1,1,1)$ ,
$u^{\mu}=g^{\mu \, \nu}\,u_{\nu}=-{1 \over \sqrt{f}}\, \delta^{\mu}_{0}$.
Here $u^{\mu}$ is a time-like four vector, $u^{\mu}\,u_{\mu}=-1. $

\noindent
The Maxwell antisymmetric tensor  and the corresponding energy momentum
tensor are respectively given by

\begin{eqnarray}
F_{\mu\, \nu}= \nabla_{\nu}\,A_{\mu}-\nabla_{\mu}\,A_{\nu}\\
M_{\mu \, \nu}={1 \over 4 \pi}\,(F_{\mu\, \alpha}\, F_{\nu}^{\alpha}-
{1 \over 4}\,F^2\,g_{\mu \, \nu})
\end{eqnarray}

\noindent
where $F^2= F^{\mu\, \nu}\,F_{\mu \, \nu}$. The current vector $j^{\mu}$
is defined as

\begin{equation}
\nabla_{\nu}\,F^{\mu\, \nu}=4\pi\,j^{\mu}
\end{equation}

\noindent
The Einstein field equations for a charged dust distribution are given by

\begin{equation}
G_{\mu\, \nu}=8\, \pi\,T_{\mu\, \nu}=8 \pi\,M_{\mu \, \nu}+
(8 \, \pi\, \rho)\, u_{\mu}\,u_{\nu}
\end{equation}

\noindent
where $\rho$ is the energy density of the charged dust distribution and the
four velocity of the dust is the same vector $u^{\mu}$ appearing in the
metric tensor. Very recently \cite{gur} we invetstigated the above filed equations
with metric given in (\ref{mt1}). We find that $j^{\mu}=\rho_{e}\, u^{\mu}$, where $\rho_{e}$
is the charge density of the dust distribution. Let $\lambda$ be a real
function depending on the space coordinates. It turns out that $A_{i}=0$ and

\begin{equation}
f={1 \over \lambda^2}, ~~A_{0}={k \over \lambda}
\end{equation}

\noindent
where $k=\pm 1$. Then the field equations reduce simply to the following
equations.

\begin{eqnarray}
\nabla^2\, \lambda +4\, \pi \, \rho\, \lambda^3=0 \label{eq2} \\
\rho_{e}=k\, \rho
\end{eqnarray}

\noindent
where $\nabla^2$ denotes the three dimensional Laplace operator in Cartesian
flat coordinates.  These equations represent the Einstein and Maxwell
equations respectively. In particular the first equation (\ref{eq2})
is a generalization of the Poisson's potential equation in
Newtonian gravity. When $\rho$ vanishes , the space-time metric
describes the Majumdar-Papapetrou space-times
 \cite{maj}\,$^{,}$\,\cite{pap}\,$^{,}$\,\cite{hah}\,$^{,}$\,\cite{ksm}.
For the case $\rho \ne 0$ , the reduced form of the field equations (\ref{eq2})
were given quite recently \cite{gur} (see also \cite{das}).

\section{Charged dust clouds}

In the Newtonian approximation $\lambda=1+V$,
Eq.(\ref{eq2}) reduces to the Poisson equation, $\nabla^2\, V+4\pi\, \rho=0$.
Hence for any physical mass density $\rho$ of the dust distribution
we solve the equation (\ref{eq2}) to find the function
$\lambda$. This determines the space-time metric completely.
As an example for a constant mass density $\rho=\rho_{0} >0$ we find that

\begin{equation}
\lambda=\displaystyle {a \over 2\, \sqrt{\pi\, \rho_{0}}}\,cn (l_{i}\, x^{i})
\end{equation}

\noindent
Here $l_{i}$ is a constant three vector ,$a^2=l_{i}l^{i}$ and $cn$ is one of
the Jacobi elliptic function with modulus square equals  ${1 \over 2}$.
This is a model universe which is filled by a (extreme) charged dust
with a constant mass density.

\section{Interior solutions}

In an asymptotically flat space-time , the function $\lambda$
asymptotically obeys the boundary condition $\lambda \rightarrow
\lambda_{0}$ (a constant).
In this case we can establish the equality of mass and charge
$e=\pm m_{0}$, where $m_{0}= \int \rho\, \sqrt{-g}\,d^3 x$. For physical
considerations our extended MP space-times
may be divided into inner and outer regions.
The interior and outer regions are defined as the regions where
$\rho_{i} >0$ and $\rho=0$ respectively. Here $i=1,2,...,N$, where
$N$ represents the number of regions. The gravitational fields of
the outer regions are described by any solution of the Laplace
equation $\nabla^2\, \lambda=0$ , for instance by the MP metrics.
As an example the extreme Reissner-Nordstr{\" o}m (RN) metric
(for $r >R_{0}$), $\lambda = \lambda_{0}+\displaystyle
{\lambda_{1} \over r}$ may be matched to a metric with

\begin{equation}
\lambda=a\, \displaystyle { \sin (b\,r) \over r},\,\, r < R_{0}
\end{equation}

\noindent
describing the gravitational field of an inner region filled by
a spherically symmetric charged dust distribution with
a mass density

\begin{equation}
\rho=\displaystyle \rho(0)\,[{b\, r \over \sin (b\,r)}]^{2}
\label{rho}
\end{equation}

\noindent
Here $\rho(0)={1 \over 4\, \pi\, a^2}$, $r^2=x_{i}\,x^{i}$,
$a$ and $b$ are constants to be determined
in terms of the radius $R_{0}$ of the boundary and total mass $m_{0}$
(or in terms of $\rho(0)$). The boundary condition , when reduced on
the function $\lambda$ on the surface $r=R$ it must satisfy both
$\lambda_{out}=\lambda_{in}$ and $\lambda^{\prime}_{out}=
\lambda^{\prime}_{in}$. Here prime denotes differentiation with respect
$r$. They lead to \cite{note}

\begin{eqnarray}
b\,R_{0}&=&\displaystyle \sqrt{ {3m_{0} \over R_{0}}},\\
\lambda_{0}&=&ab \cos (bR_{0}),\\
\lambda_{1}&=&a\,(\sin (bR_{0}) -bR_{0}\, \cos (bR_{0}))
\end{eqnarray}

\noindent
When the coordinates transformed to the Schwarzschild coordinates
.i.e.,  $ \lambda_{0}\,r+\lambda_{1} \rightarrow r $ then the line element
becomes

\begin{equation}
ds^2=\displaystyle -{1 \over \lambda_{0}^2}(1- {\lambda_{1} \over r})^2\, dt^2+
{dr ^2 \over( 1-{\lambda_{1} \over r})^2}+r^2\,d \Omega^2.
\end{equation}

\noindent
hence $\lambda_{1}$ is the mass in the Newtonian approximation then
$\lambda_{1}=m_{0}$.

\noindent
In this way one may eliminate
the singularities of the outer solutions by matching them to an inner
solution with a physical mass density.

For the  mass density
$\rho= {b^{2} \over  4 \pi\,\lambda^2}$ in general ,
we may have the complete solution.
Here $b$ is a nonzero constant which is related to $m_{0}$ by the relation
$b^2\,R_{0}^3=3 m_{0}$ and we find that

\begin{equation}
\lambda=\displaystyle \sum_{l,m}\,a_{l,m}\,
j_{l}(b\,r)\,Y_{l,m}(\theta, \phi)
\end{equation}

\noindent
where $j_{l}(b\,r)$ are the spherical Bessel functions which are
given by

\begin{equation}
j_{l}(x)=\displaystyle (-x)^l\,\big ( {1 \over x}\, {d \over dx} \big )^l\,
\big ( {\sin x \over x} \big )
\end{equation}

\noindent
and $Y_{l,m}$ are the spherical harmonics. The constants
$a_{l,m}$ are determined when this solution is matched to an outer solution
with $\nabla^2 \lambda=0$. The interior solution given above for
the extreme RN metric with density (\ref{rho}) corresponds to $l=0$.

\section{Point particle solutions}

Newtonian gravitation is governed by the Poisson type
of linear equation. Gravitational fields of spherical objects
in the exterior regions may also be identified as the gravitational
fields of masses located at a discrete points (point particles located
at centers of the spheres) in space $(R^3)$. The solution
of the Poisson equation with $N$ point singularities may be given by

\begin{eqnarray}
\lambda= 1+\displaystyle \sum_{i=1}^{N} {m_{i} \over r_{i}},\label{pnt}\\
r_{i}=[(x-x_{i})^2+(y-y_{i})^2+(z-z_{i})^2]^{1 \over 2}
\end{eqnarray}

\noindent
where $N$ point particles with masses $m_{i}$ are located at the points
$(x^{i},y^{i},z^{i})$ with $i=1,2,,,N$. The same solution given above
may also describe the exterior solution of the $N$ spherical objects
(with nonempty interiors) with total masses $m_{i}$ and radii $R_{i}$.
The interior gravitational fields of such  spherical objects can be
determined when the mass densities $\rho_{i}$ are given. The essential point
here is that the limit $R_{i} \rightarrow 0$  is allowed. This means that
the dust distribution is replaced by a distribution concentrated at the
points $(x^{i},y^{i},z^{i})$  .
Namely in this limit mass densities behave as the Dirac delta functions,

$$ \displaystyle \rho \rightarrow \sum_{i=1}^{N} m_{i}\, \delta (x-x_{i})\,
\delta (y-y_{i})\, \delta (z-z_{i}).$$

\noindent
This is consistent with the
Poisson equation $\nabla^2 \lambda +4\pi\, \rho\,=0$ , because
${1 \over r_{i}}$ in the solution (\ref{pnt}) is the Green's function, i.e.,

$$ \displaystyle \nabla^2 {1 \over r_{i}}= -4\pi\,  \delta (x-x_{i})\,
\delta (y-y_{i})\, \delta (z-z_{i}).$$

Such a limit , i.e., $R_{i} \rightarrow 0$ is not consistent
in our case ,  $\nabla^2 \lambda +4\pi\, \rho\, \lambda^3=0$.
The potential equation is nonlinear and in particular in this limit
the product of $\rho$ and $\lambda^3$ does not make sense. Hence we
remark that the Majumdar-Papapetrou metrics should represent
the gravitational field $N$ objects with nonempty interiors
(not point-like objects).

\section{Thin shell solutions}

In the previous section we concluded that the dust distribution can
not be concentrated to a point. We observe that ,
the potential equation (\ref{eq2}) does not also admit dust distributions on
 one dimensional (string like distributions) structures.
This is compatible with the results of Geroch-Traschen \cite{ger}.
On the other hand , the mass distribution $\rho$ can be defined on surfaces.

Let $S$ be a regular surface in space ($R^3$) defined by
$S=[(x,y,z) \in R^3 ; F(x,y,z)=0]$ , where $F$ is a differentiable
function in $R^3$. When the dust distribution is concentrated on
$S$ the mass density may be represented by the Dirac delta
function

\begin{equation}
\rho (x,y,z)= \rho_{0}(x,y,z)\, \delta (F(x,y,z))
\end{equation}

\noindent
where $\rho_{0}(x,y,z)$ is a function of $(x,y,z)$ which is defined on $S$.
The function $\lambda$ satisfying the potential equation (\ref{eq2})
compatible with such shell like distributions may given as

\begin{equation}
\lambda (x,y,z)=\lambda_{0} (x,y,z)-\lambda_{1} (x,y,z)\, \theta (F)
\end{equation}

\noindent
where $\lambda_{0}$ and $\lambda_{1}$ are differentiable functions of $(x,y,z)$
and $\theta (F)$ is the Heaviside step function. With these assumptions
we obtain

\begin{eqnarray}
\rho_{0}(x,y,z)={1 \over 4 \pi}\,{\vec{\nabla} \lambda_{1} \cdot \vec
{\nabla} F \over (\lambda_{0} )^3} \vert_{S}\\
\nabla^2 \lambda_{0}= \nabla^2 \lambda_{1}=0
\end{eqnarray}

\noindent
and in addition $\lambda_{1}\vert_{S}=0$. We have some examples:

\vspace{0.5cm}

\noindent
{\bf 1.}\,\, $S$ is the plane $z=0$. We have $\rho (x,y,z)=
 \rho_{0}(x,y)\, \delta (z)$. Then it follows that

\begin{eqnarray}
\lambda (x,y,z)&=&\lambda_{0}(x,y,z)-\lambda_{2} (x,y)\,z\, \theta (z)\\
\rho_{0} (x,y,z)&=&{1 \over 4 \pi}\,{ \lambda_{2} \over (\lambda_{0} \vert_{S})^3}\\
\nabla^2 \lambda_{0}&=& \nabla^2 \lambda_{2}=0
\end{eqnarray}

\vspace{0.5cm}

\noindent
{\bf 2.}\,\, $S$ is the cylinder $F=r-a=0$. We have $\rho (r,\theta,z)=
 \rho_{0}(\theta,z)\, \delta (r-a)$. Then it follows that

\begin{eqnarray}
\lambda (r,\theta,z)&=&\lambda_{0}(r,\theta,z)-\lambda_{2}
(\theta,z)\,\ln (r/a)\, \theta (\rho-a)\\
\rho_{0} (r,\theta,z)&=&{1 \over 4 \pi a}\,{ \lambda_{2} \over (\lambda_{0} \vert_{S})^3}\\
\nabla^2 \lambda_{0}&=& \nabla^2 \lambda_{2}=0
\end{eqnarray}

\noindent
Here we remark that the limit $a \rightarrow 0$ does not exist.
This means that the mass distribution on the whole $z$- axes is not allowed.

\vspace{0.5cm}

\noindent
{\bf 3.}\,\, $S$ is the sphere $F=r-a=0$. We have $\rho (r,\theta,\phi)=
 \rho_{0}(\theta,\phi)\, \delta (r-a)$. Then it follows that

\begin{eqnarray}
\lambda (r,\theta,\phi)&=&\lambda_{0}(r,\theta,\phi)-\lambda_{2}
(\theta,\phi)\,({1 \over a}-{1 \over r})\, \theta (r-a)\\
\rho_{0} (r,\theta, \phi)&=&{1 \over 4 \pi a^2}\,{ \lambda_{2} \over (\lambda_{0}
\vert_{S})^3}\\
\nabla^2 \lambda_{0}&=& \nabla^2 \lambda_{2}=0
\end{eqnarray}

\noindent
We note that the total mass is infinite on non-compact surfaces.

For compact case we shall consider the sphere in more detail.
In this case we may have $\lambda_{0}= \mu\,\lambda_{2}+\psi$ such that
$\nabla^2 \psi =0$ and $\psi (a, \theta, \phi)=0$. We shall assume $\psi=0$
everywhere, then

\begin{equation}
\rho_{0}= {1 \over 4 \pi a^2}\, {1 \over \mu^3\,\lambda_{2}^2}
\end{equation}

\noindent
Hence the total mass $m_{0}$ on $S$ is given by
$m_{0}=\int \sqrt{-g}\, \rho\,d^3\,x={1 \over \mu}$. Let
$\lambda^{out}$ and $\lambda^{in}$ denote solutions
of (\ref{eq2}) corresponding to the exterior and inner regions respectively.
They are given by

\begin{eqnarray}
\lambda^{out} (r,\theta,\phi)&=&\lambda(r>a, \theta,\phi)=1 -{m_{0} \over a}+
{m_{0} \over r}\, \label{sl1}\\
\lambda^{in} (r,\theta,\phi)&=&\lambda(r<a,\theta,\phi)=1 \label{sl3},
\end{eqnarray}

\noindent
where we let $\lambda_{2}=m_{0}$. Here we remark that the point particle limit
$a \rightarrow 0$ does not exist. The solution given above represents the
extreme Reissner-Nordstr{\" o}m solution

\begin{equation}
ds^2=-\,(1 -{m_{0} \over a}+{m_{0} \over r})^{-2} dt^2+
(1 -{m_{0} \over a}+{m_{0} \over r})^2\,(dr^2+r^2\, d\, \Omega^2)
\end{equation}

\noindent
By letting $r={R-m_{0} \over \beta}$ where $\beta=1-{m_{0} \over a}$
and $a \ne m_{0}$ is assumed.
We obtain the extreme RN in its usual form

\begin{equation}
ds^2=-\,\beta^2\,(1 -{m_{0} \over R})^{2} dt^2+
{dR^2 \over (1 -{m_{0} \over R})^2}+R^2\, d\, \Omega^2
\end{equation}

\noindent
Hence we obtain a solution where
the exterior solution is the extreme
Reissner - Nordstr{\" o}m metric, but inside the sphere with radius
$a$, the spacetime is flat. Thus extreme RN solution is matched to a
spherical shell ($R=a$) of dust distrubition.

The case $a=m_{0}$ represents the
Levi-Civita -Bertotti-Robinson spacetime outside the dust shell and the flat
spacetime inside.

\begin{equation}
ds^2=-{r^2 \over m_{0}^2}\,dt^2+ {m_{0}^2 \over r^2}\,dr^2+
m_{0}^2\, d\, \Omega^2 \label{br}
\end{equation}

\noindent
By letting $r={ m_{0}^2 \over R}$ we obtain the usual conformally flat
$LCBR$ metric

\begin{equation}
ds^2={m_{0}^2 \over R^2}\, [-dt^2+dR^2+R^2\, d \Omega^2]
\end{equation}

\noindent
In the new coordinates the surface is again $R=m_{0}$. In these dust shell
solutions the function $\lambda$ is continuous on the surface $r=a$ , but
its normal derivative to $S$ is discontinuous ,

$$\lambda^{\prime}_{out}
-\lambda^{\prime}_{in}=-4 \pi\,\sigma=-m_{0}/a^2$$

\noindent
as expected, where $\sigma=$ mass per unit area $=m_{0}/4\pi a^2$.
For thin shells in general relativity see the recent work
of Mansouri and Khorrami \cite{man} and also Mansouri's contribution
in this proceeding.

\section{Conclusion}

We have solved the Einstein field equations in a
conformo-static space-time for a charged dust distribution. We
reduced the whole Einstein field equations to a nonlinear Poisson
type of potential equation (\ref{eq2}). Physically
reasonable solutions of this equation
give an interior solution to an exterior MP metrics.
We have given some explicit exact solutions corresponding to some
mass densities. In particular we have given an interior solution of the
extreme RN metric.

We showed that the limiting cases of mass distributions on discrete points
and also on lines in $R^3$ are not possible.
We have examined some possible mass distributions on regular
surfaces. We have found solutions corresponding to shell like dust
distributions. In particular for the spherical dust shell
we presented an exact solution given in eqns. (\ref{sl1}-\ref{br})
representing a spherical shell cavity immersed in the extreme RN spacetime.

\section*{Acknowledgments}

This work is partially supported by the Alexander von Humboldt
Foundation and Turkish Academy of Sciences.

\end{document}